# Photo-Induced Bandgap Renormalization Governs the Ultrafast Response of Single-Layer MoS$_2$

*Eva A. A. Pogna*[1,†], *Margherita Marsili*[2,†], *Domenico De Fazio*[3], *Stefano Dal Conte*[1,4], *Cristian Manzoni*[1,4], *Davide Sangalli*[5], *Duhee Yoon*[3], *Antonio Lombardo*[3], *Andrea C. Ferrari*[3], *Andrea Marini*[5,*], *Giulio Cerullo*[1,4,*], *Deborah Prezzi*[2,*]

[1]Dipartimento di Fisica - Politecnico di Milano, Piazza Leonardo da Vinci 32, I-20133 Milano (IT)

[2]Centro S3, Istituto Nanoscience (NANO) - CNR, via Campi 213/a, I-41125, Modena (IT)

[3]Cambridge Graphene Centre, University of Cambridge, Cambridge CB3 OFA (UK)

[4]Istituto di Fotonica e Nanotecnologie (IFN) - CNR, I-20133 Milano (IT)

[5]Istituto di Struttura della Materia (ISM) - CNR, Via Salaria Km 29.3, Monterotondo Stazione (IT)

† These authors contributed equally to the work.

* Corresponding authors: (GC) giulio.cerullo@polimi.it; (AM) andrea.marini@cnr.it; (DP) deborah.prezzi@nano.cnr.it

ABSTRACT. Transition metal dichalcogenides (TMDs) are emerging as promising two-dimensional (2d) semiconductors for optoelectronic and flexible devices. However, a microscopic explanation of their photophysics – of pivotal importance for the understanding and optimization of device operation – is still lacking. Here we use femtosecond transient absorption spectroscopy, with pump pulse tunability and broadband probing, to monitor the relaxation dynamics of single-layer $MoS_2$ over the entire visible range, upon photoexcitation of different excitonic transitions. We find that, irrespective of excitation photon energy, the transient absorption spectrum shows the simultaneous bleaching of all excitonic transitions and corresponding red-shifted photoinduced absorption bands. First-principle modeling of the ultrafast optical response reveals that a transient bandgap renormalization, caused by the presence of photo-excited carriers, is primarily responsible for the observed features. Our results demonstrate the strong impact of many-body effects in the transient optical response of TMDs even in the low-excitation-density regime.

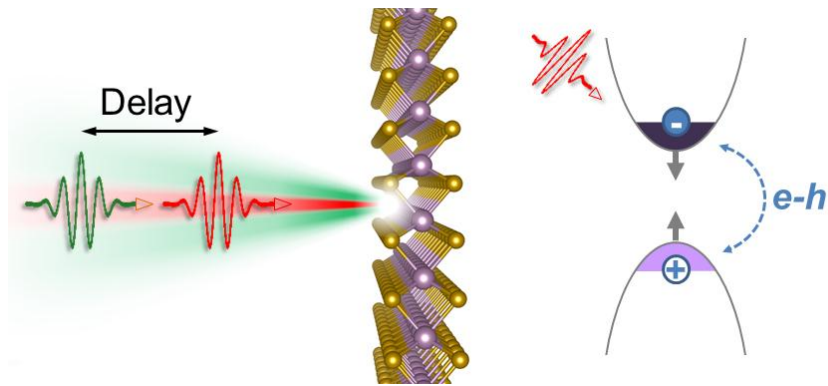

Mo- and W-based dichalcogenides are semiconducting layered crystals that are attracting an ever increasing attention for the potential technological impact of their unique electronic and optical properties.[1–5] In the single-layer (1L) limit, they exhibit an indirect-to-direct band-gap transition, which is accompanied by efficient light emission in the NIR-visible range.[6,7] In addition, the coupling of spin and valley degrees of freedom gives rise to valley-selective optical properties.[8–10] Interestingly, markedly enhanced many-body Coulomb interactions emerge in these atomically thin materials due to the extreme quantum confinement and reduced screening. This yields large exciton binding energies of several hundreds of meV,[11–13] together with other prominent multi-particle excitations such as trions,[14–16] and biexcitons.[17]

The recent attempts to harness these properties for new-generation optoelectronic devices – from photo-detectors[18] to light-emitting diodes–,[19] have triggered an intense research activity devoted to the fundamental understanding of the photo-physics of these materials, with special focus on $MoS_2$. A number of time-resolved studies have shown that, despite differences related to the nature of the samples, some common features are ubiquitously observed, both in 1L- and few-layer $MoS_2$[20–26] as well as in other TMDs:[27,28] The transient absorption (TA) spectra always exhibit a bleaching of the lowest-energy excitonic absorption features, accompanied by red-shifted photo-induced absorption sidebands. The interpretation is however controversial, ranging from carrier-induced broadening,[22] to exciton-biexciton transitions.[26] On the theory side, the investigation of the carrier dynamics is only at the beginning.[24,29,30]

In this work, we assign these features by combining ultrafast TA spectroscopy of exfoliated 1L-$MoS_2$, with selective excitation of the three excitonic resonances and broadband probing, and time-domain first-principles simulations, accounting for electronic correlations and spin-orbit effects. This approach allows us to establish the dominant role of renormalization effects, stemming from the partial compensation of electronic band-gap shrinkage and exciton binding energy reduction induced by the photo-excited carriers.

RESULTS AND DISCUSSION

Experiments are performed on a 1L-MoS$_2$ flake, prepared by micromechanical cleavage of the bulk crystal and transferred by a wet-transfer technique onto a fused silica substrate,[4,5] allowing a transmission geometry (see Figure 1a). The samples are characterized by photoluminescence and Raman spectroscopy, both prior and after transfer (see Methods). The static absorption spectrum of 1L-MoS$_2$ is also measured, as discussed in Methods, and is reported in Figure 1b. The spectrum is characterized by three main peaks of excitonic nature at ~1.9, 2.04 and 2.9 eV, the so-called A, B and C excitons, respectively, in agreement with previous observations.[6,7,31] The A and B excitons result from optical transitions between the spin-orbit-split top valence band and the bottom conduction band, around K and K';[31–33] the C resonance is ascribed to a weakly bound exciton, with a complex composition mainly resulting from interband transitions around Γ.[31,34]

The transient optical response is measured with a custom-built broadband TA microscope (see Figure 1a), where a tunable pump pulse is combined with broadband white-light probe to span the entire visible range, thus covering the three main excitonic transitions (see Methods for a detailed description). Figure 1c shows a 2d TA map as a function of probe photon energy and pump-probe delay, for a pump photon energy of 3.1 eV, *i.e.* above the C peak and the electronic band gap of 1L-MoS$_2$ (~2.8 eV).[35] For each delay, the TA spectrum exhibits three prominent features in correspondence of the A, B and C excitons. Each feature consists of a reduced absorption (photo-bleaching) at the excitonic resonance (negative signal) and a corresponding red-shifted photo-induced absorption (positive signal). The relative intensity of these peaks changes for time delays up to 100 ps, but no further spectral components appear in this temporal range.

In order to probe specific features of different excitonic transitions, the transient optical response of 1L-MoS$_2$ is investigated also as a function of the excitation energy. Figure 2 compares, for a fixed delay of 300 fs, the TA spectrum for 3.1 eV pump photon energy to those recorded with excitation in resonance with the A (1.88 eV) and B (2.06 eV) excitons. Despite the different excitation conditions, the three TA spectra bear a remarkable similarity, and each of them shows the previously discussed negative (photo-bleaching) and positive (photo-induced absorption) features in correspondence of A, B and C, although the relative intensity of these peaks changes with the pump photon energy. Similar features were previously observed in the TA spectra of 1L- and few-layer MoS$_2$,[20–26] as well as in other TMDs[27,28] – even though probing was limited to the low-energy region (A and B peaks) –, with a variety of interpretations: The absorption reduction at the excitonic resonances has been alternatively attributed to Pauli blocking,[20,21,23] stimulated emission,[24,26] or exciton self-energy renormalization.[22] The photo-induced absorption, instead, has been explained invoking carrier-induced broadening[21,22,24] or biexciton formation.[26] Our experimental evidence, with tunable excitation and broadband probing, puts these interpretations in a different perspective. We observe a simultaneous change in the system response throughout the probed energy range, regardless of excitation photon energy. This shows a resolution-limited build-up and appears already in the low-excitation regime, where contributions from multi-particle excitations are negligible. This suggests a more general electronic mechanism, in action from the very early stages of the carrier dynamics.

In order to identify the underlying process, we perform time-domain *ab-initio* simulations[36–39] that combine non-equilibrium Green's function and density-functional theories, as implemented in the Yambo package[40] (see Methods for a detailed description). Within this approach, we compute from first principles both the static absorption spectrum, within the *GW* plus Bethe-Salpeter (BS) scheme,[41] and the TA spectrum, following the system excitation with an ultra-short pump pulse. This method reproduces the experimental excitation conditions, and fully takes into account electron-hole (e-h) interactions, both in the static optical response and in the non-equilibrium dynamics. This is of fundamental importance when investigating TMDs and other low-dimensional materials, where Coulomb interactions are known to dominate the optical response.[11,12] Spin-orbit coupling is also taken into account to correctly describe the A-B splitting.[6,7]

Figure 3 plots the simulated TA spectrum. As in the experiments, we consider three different excitation pulses, with photon energies in resonance with A and B, and 0.1 eV above the C peak of the theoretical absorption spectrum, the latter mimicking the off-resonance condition. Starting from the non-equilibrium carrier population created by the pump pulse, we compute the perturbed optical response of the system. Since we are in a regime of low-carrier concentration, one option would be to assume that the excitation energies (*i.e.* the poles of the response function) of the perturbed system are the same as those at equilibrium. This corresponds to interpreting the observed features as the result of a change in the electronic occupations only, *i.e.* Pauli blocking. However, the TA spectra computed within this approximation (Figure 3a) do not reproduce the experimental results, as they mainly show a bleaching of the exciton that is in resonance with, or closest to, the excitation frequency. As a consequence of Pauli blocking, a simultaneous bleaching of the three excitons can only appear if they are composed by the same e-h pair population. This is not the case here, as demonstrated by the exciton composition analysis (see Supporting Information, Figure S2), in agreement with previous calculations.[31] Thus, the simultaneous bleaching cannot be interpreted solely in terms of Pauli blocking.

However, photo-generation of excited-state carriers may also alter the excitons' energy and composition by affecting the underlying electronic states. This occurs because the modification of the electronic occupations provides both additional screening channels and charge density variations. Figure 3b displays the TA spectra obtained by including all the modifications induced by the photo-excited carrier population. This shows a striking effect on the TA spectra, with a much better agreement with the experiments, compared to the case where Pauli blocking only is included (Figure 3a). In all pump conditions, we find a simultaneous bleaching of the three excitonic transitions. In addition, positive features appear in correspondence of the peaks, red-shifted with respect to the exciton energies. As schematically depicted in Figure 4, the enhancement of the electronic screening leads to the simultaneous renormalization of both binding energy and band-gap (BGR), giving rise to opposite effects that partially compensate each other: a red-shift of the optical spectra due to electronic gap shrinkage, and a blue-shift arising from binding energy reduction. Even though the quantitative detail of this compensation depend on both material and dimensionality, this is a general effect in the low carrier-density regime,[42] studied in several systems from 3d semiconductors[43] to 1d quantum wires.[44] Here the compensation is only partial and gives rise to an overall peak shift of a few tens meVs, in agreement with recent photoluminescence and TA data.[45,46] This explains the characteristic derivative-like lineshape of the TA spectra, with alternating positive and negative peaks at the excitonic resonances, as observed experimentally. Note also that, since this renormalization is due to a change of the potential felt by electrons and holes, it affects all states simultaneously and independently of their energy, again as observed experimentally in the present work. Different phenomena arise instead in the high carrier-density regime, where the dynamics is dominated by the presence of a hot e-h plasma, as predicted theoretically for $MoS_2$[29] and recently observed in $WS_2$.[47]

CONCLUSIONS.

By probing the transient optical response of 1L- $MoS_2$ with unprecedented spectral coverage and tunable pump, we demonstrate that exciton photobleaching and red-shifted photo-induced absorption are not limited to the lowest-energy excitation, but appear throughout the probed energy range, independently of excitation energy. The application of a novel ab-initio approach, able to describe the carrier dynamics upon photoexcitation by fully accounting for many-body effects, allows us to reveal that the mechanism responsible for the simultaneous bleaching of the three excitonic peaks and the corresponding red-shifted photo-induced absorption is the renormalization of the excitation energies in presence of photo-excited carriers. Our results provide a global understanding of the ultrafast photophysics of $MoS_2$, in the early stages after photoexcitation, in terms of fundamental phenomena, which are expected to govern the physics of the entire class of 2d TMDs.

METHODS.

SAMPLE PREPARATION AND CHARACTERIZATION. $MoS_2$ flakes are prepared by micromechanical cleavage of bulk crystal of natural molybdenite (Structure Probe Inc., SPI) on a Si wafer covered with 285 nm of $SiO_2$. 1L-$MoS_2$ flakes were then identified by optical contrast, photoluminescence and Raman spectroscopy measurements.[48,49] The selected 1L-$MoS_2$ flakes are then moved onto 100 μm fused silica substrates (suitable to perform TA in transmission in view of its isotropy) by a wet-transfer technique based on a sacrificial layer of poly-methyl-methacrylate (PMMA).[5] To easily locate the samples for TA measurements, a metal frame is fabricated around selected flakes by photolithography, followed by thermal evaporation of 2-nm Cr and 100-nm Au films. Photoluminescence and Raman spectroscopy measurements are repeated after the transfer and the metal frame deposition, showing no significant changes.[50]

The sample optical absorption is measured using a broadband white light from a tungsten halogen lamp. The transmitted light is collected with a 100x objective (0.9 numerical aperture). A 50μm pinhole in a Horiba LabRAM HR800 spectrometer is used to achieve a~2μm resolution. The collected transmitted light is dispersed by a 150 grooves/mm grating and measured with a liquid nitrogen cooled charge coupled device detector. The absorbance is defined as $(T_0-T)/T_0$, where $T_0$ is the transmitted light through the quartz substrate and T the transmitted light through 1L-$MoS_2$ on quartz substrate.

PUMP-PROBE SETUP. We use an amplified Ti:Sapphire laser (Coherent Libra), emitting 100-fs pulses at 1.55 eV, with average power of 4 W at 2 kHz repetition rate. A 320-mW fraction of the laser power is used for our experiments. The output beam is divided by a 50/50 beam splitter into pump and probe lines. We produce ~100-fs pump pulses at 3.1 eV by second harmonic generation, while we use an optical parametric amplifier to generate ~70-fs pump pulses at ~2 and 1.9 eV. The probe is obtained by white light continuum generation in a 5-mm-thick $CaF_2$ plate, filtering out light at the fundamental frequency with a colored glass short-pass filter, and it shows rms fluctuations as low as 0.2%. The collinear pump and probe pulses are focused by a reflective microscope objective with 15x magnification to avoid chromatic aberrations. At the sample position, the pump and probe spots have a diameter of 10 and 5 μm, respectively, as determined by knife-edge measurement using the edge of a pinhole on the metal mask. The transmitted probe is collected with a 7.5 mm achromatic doublet, dispersed by a $CaF_2$ prism and detected by a Si CCD camera with 532 pixels, corresponding to a bandwidth per pixel of 1.1 nm. The pump and probe pulses have perpendicular polarizations and a linear polarizer is used to filter out the pump light scattered from the sample. The pump pulse is modulated at 1 kHz by a mechanical chopper such that a differential signal of the order of $10^{-4}$ could be detected with an integration time of 2 s. For our setup, we estimate an overall temporal resolution of ~150fs. Although the probe pulse is not transform-limited, in the case of broadband multichannel detection the effective temporal resolution is almost the same as with a transform-limited probe pulse.[51,52] The pump fluence is ~10μJ/cm$^2$, well below the damage threshold of 50μJ/cm$^2$.[53] This corresponds to a photogenerated carrier density of 2-3×10$^{12}$ cm$^{-2}$ (depending on the pump photon energy), well-below the Mott density measured for this material.[54] The initial carrier density, calculated assuming that each absorbed photon generates one carrier,[23] is determined from the experimental absorption of 1L-MoS$_2$ at each pump photon energy (Figure 1b), taking into account the light scattering background (5.7% at 1.88 eV, 6.1 % at 2.06 eV, and 15% at 3.1 eV). A detailed scheme of the experimental apparatus is reported in Figure S1 of Supporting Information.

COMPUTATIONAL DETAILS. We perform density-functional theory (DFT) calculations, within the local-density approximation (LDA), using the Quantum Espresso package.[55] We employ fully relativistic norm-conserving pseudopotentials, including for Mo semicore 4p, 4d and 4s states in the valence band. The computed equilibrium lattice constant is ~3.12 Å at 0 K, in good agreement with the experimental lattice constant, 3.13 Å at 77 K.[56] The equilibrium absorption spectrum is calculated by solving the BS equation for the four-point polarizability on top of the GW-corrected band structure,[41] as implemented in the Yambo package.[40] A distance of 21 Å is kept in the out of plane direction, together with the application of a truncation scheme for the Coulomb potential,[57] to avoid spurious interactions between sheet replicas. The irreducible Brillouin zone (IBZ) is sampled with a 24×24×1 *k*-point grid centered at Γ. The random-phase approximation (RPA) polarizability is built using a cutoff of 8 Ry in the reciprocal space, and including in the summation states up to 30 eV above the conduction band minimum. The BS kernel is computed using two occupied and two empty bands. This minimal set is chosen to guarantee the feasibility of real-time simulations, while ensuring a good description of the main excitonic features. A scissor operator of -0.2 eV is further applied to match the experimental position of the first excitonic peak, and an energy-dependent broadening is added to mimic lifetime effects[31] (see Figure S2 of Supporting Information).

The non-equilibrium population of the electronic states in the presence of the pump laser pulse is obtained by following the time evolution of the lesser Green's function ($G^<$) with the static Coulomb-hole screened exchange expression (CoHSEx) for the electronic self-energy,[41] as described in Refs. 36–39. The polarization created by the laser pulse is then dephased with a constant (in time) term of 20 meV, consistent with the broadening used for the absorption at equilibrium. We do not consider the time evolution of the occupations due to scattering events, since the goal of our simulations is to describe the origin of the simultaneous bleaching of the excitons, which appears immediately after the pump and not during the time evolution. The laser pulses are simulated as a sinusoidal time-dependent external potential convoluted with a Gaussian function with 100 fs standard deviation, and intensity 2000 kW/m$^2$. This value of the field, corresponding to a carrier density of 0.6-1 x 10$^{11}$ cm$^{-2}$ (depending on the pump photon energy), gives TA spectra quantitatively comparable to those measured experimentally.

Moreover, we have checked to be in the linear regime for the chosen value of the field (density), as in the experimental case. This holds for simulated carrier densities up to $10^{12}$ cm$^{-2}$, giving a behavior consistent with the experiments regardless of the exact value used in the simulations. Instead, the simulated carrier density should not be directly compared with that estimated experimentally, as the evaluation of this quantity relies on several assumptions in both theory and experiments.

FIGURES

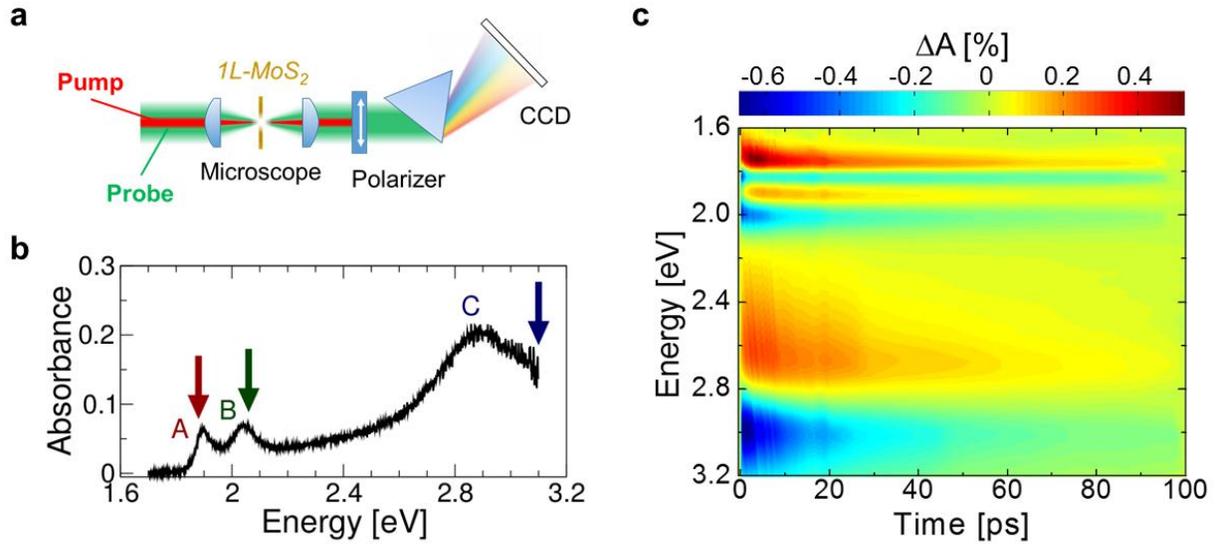

**Figure 1.** (a) Schematic of the pump-probe microscope. A femtosecond pump pulse excites 1L-MoS$_2$, while a delayed broadband probe pulse monitors the changes in the optical transmission spectrum as function of delay. (b) Absorbance of 1L-MoS$_2$. The pump-probe experiment is performed by pumping in resonance with A ($\omega_{pump}$ = 1.88 eV, red arrow) and B ($\omega_{pump}$ = 2.06 eV, green arrow), and out of resonance with C ($\omega_{pump}$ = 3.10 eV, blue arrow); the probe pulse covers the entire visible range (1.6-3.2 eV). (c) Transient absorption map as a function of probe photon energy and pump-probe delay, after photoexcitation at $\omega_{pump}$ = 3.10 eV.

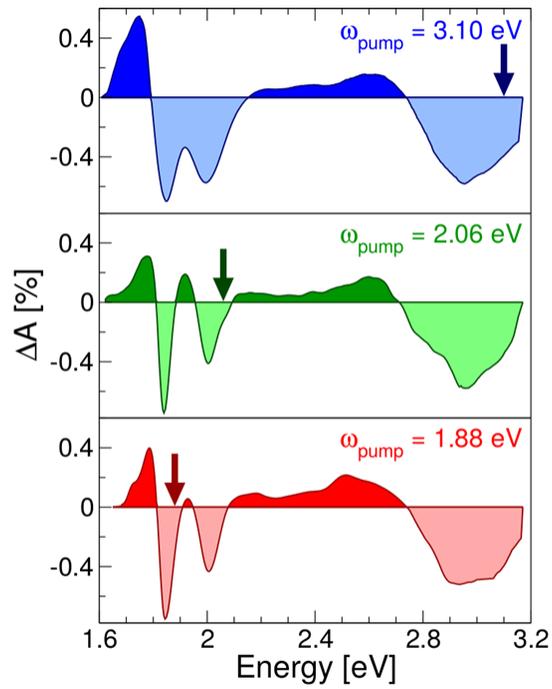

**Figure 2.** Transient absorption spectra of 1L-MoS$_2$, recorded at fixed pump-probe delay $t$ = 300 fs, for three pump photon energies, *i.e.* in resonance with the A ($\omega_{pump}$ = 1.88 eV) and B ($\omega_{pump}$ = 2.06 eV) excitons, and out of resonance with C ($\omega_{pump}$ = 3.10 eV).

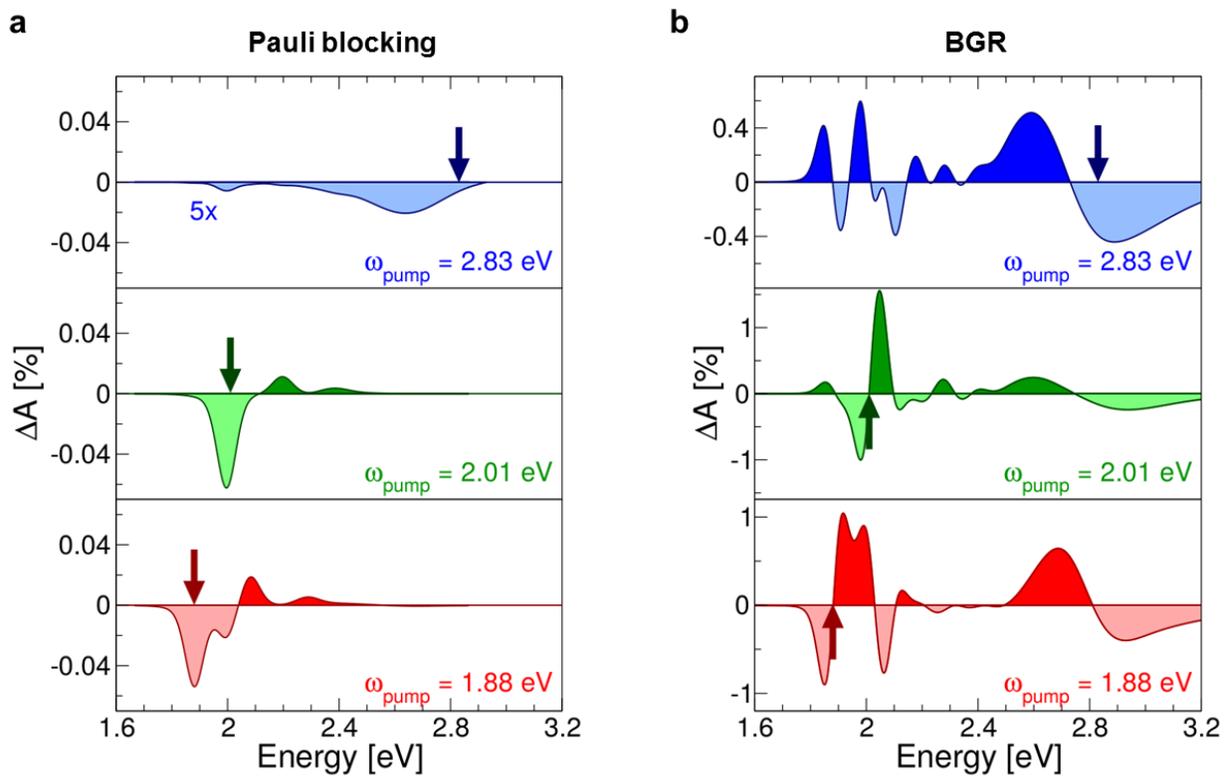

**Figure 3.** Calculated transient absorption spectra of 1L-MoS$_2$ for three pump photon energies, *i.e.* 1.88 eV, 2.01 eV, and 2.83 eV. The first two energies correspond to excitation of the system in resonance with the A and B excitons; the higher energy is 0.1 eV above the theoretical position of the C peak. The results obtained by including Pauli blocking only (a) are compared to those obtained including also the effect of the renormalization of electronic gap and excitonic binding energies (b).

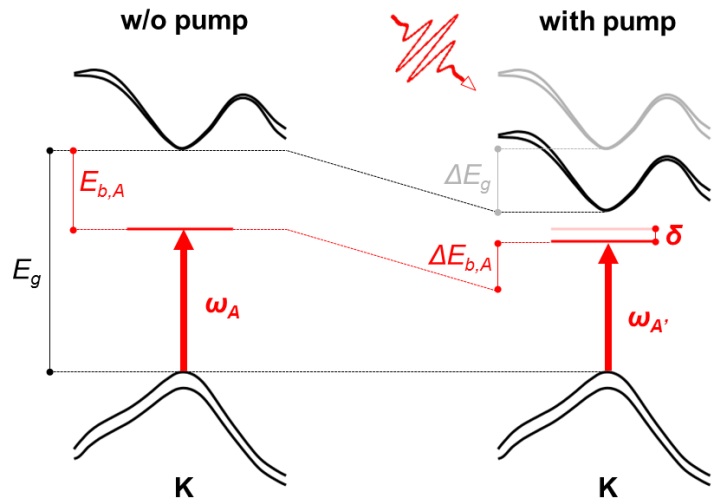

**Figure 4.** Electronic band structure (black) of 1L-MoS$_2$ at the K point and its optical excitations (red) in absence (left) and in presence (right) of pump pulse absorption. The overall shift of the excitonic absorption resonances ($\delta$) results from both the shrinkage of the electronic band gap and the renormalization of the exciton binding energy, due to the presence of photo-excited carriers, *i.e.,* $\delta = \Delta E_g - \Delta E_b$.

## ASSOCIATED CONTENT

**Supporting Information**. A detailed scheme of the experimental apparatus, the calculated absorption spectrum and exciton composition analysis are available as Supporting Information. This material is available free of charge *via* the Internet at http://pubs.acs.org.

## AUTHOR INFORMATION

**Corresponding Author**

* Correspondence should be addressed to: (GC) giulio.cerullo@polimi.it; (AM) andrea.marini@cnr.it; (DP) deborah.prezzi@nano.cnr.it

**Author Contributions**

G.C., D.P. e A.M. conceived the work. D.D.F., M.B., I.G. and A.C.F were responsible for the sample preparation and characterization. E.A.A.P., C.M and S.D.C. carried out the pump-probe measurements. M.M. performed the ab-initio simulations. A.M. and D.S. implemented the theoretical and methodological tools used for the calculations. D.P., A.M. and G.C. wrote the paper with contributions from all authors. All authors discussed the results and commented on the manuscript at all stages, and have given approval to the final version of the manuscript.

† These authors contributed equally to the work.

## ACKNOWLEDGMENTS

We acknowledge funding from MIUR FIRB Grant No. RBFR12SWOJ, MIUR PRIN Grant No. 20105ZZTSE, MAE Grant No. US14GR12, EU Graphene Flagship (Contract No. CNECT-ICT-604391), ERC Synergy Hetero2D, a Royal Society Wolfson Research Merit Award, EPSRC grants EP/K01711X/1, EP/K017144/1, EP/L016087/1. Computing time was provided by the Center for Functional Nanomaterials at Brookhaven National Laboratory (supported by the U.S. Department of Energy, Office of Basic Energy Sciences, under contract number DE-SC0012704), and by CRESCO/ENEAGRID High Performance Computing infrastructure (funded by the Italian National Agency for New Technologies, Energy and Sustainable Economic Development ENEA and by Italian and European research programmes).

REFERENCES

(1) Wang, Q. H.; Kalantar-Zadeh, K.; Kis, A.; Coleman, J. N.; Strano, M. S. Electronics and Optoelectronics of Two-Dimensional Transition Metal Dichalcogenides. *Nat. Nanotechnol.* **2012**, *7*, 699–712.

(2) Butler, S. Z.; Hollen, S. M.; Cao, L.; Cui, Y.; Gupta, J. A.; Gutiérrez, H. R.; Heinz, T. F.; Hong, S. S.; Huang, J.; Ismach, A. F.; Johnston-Halperin, E.; Kuno, M.; Plashnitsa, V. V.; Robinson, R. D.; Ruoff, R. S.; Salahuddin, S.; Shan, J.; Shi, L.; Spencer, M. G.; Terrones, M.; *et al.* Progress, Challenges, and Opportunities in Two-Dimensional Materials beyond Graphene. *ACS Nano* **2013**, *7*, 2898–2926.

(3) Sundaram, R. S.; Engel, M.; Lombardo, A.; Krupke, R.; Ferrari, A. C.; Avouris, P.; Steiner, M. Electroluminescence in Single Layer MoS2. *Nano Lett.* **2013**, *13*, 1416–1421.

(4) Ferrari, A. C.; Bonaccorso, F.; Fal'ko, V.; Novoselov, K. S.; Roche, S.; Bøggild, P.; Borini, S.; Koppens, F. H. L.; Palermo, V.; Pugno, N.; Garrido, J. A.; Sordan, R.; Bianco, A.; Ballerini, L.; Prato, M.; Lidorikis, E.; Kivioja, J.; Marinelli, C.; Ryhänen, T.; Morpurgo, A.; *et al.* Science and Technology Roadmap for Graphene, Related Two-Dimensional Crystals, and Hybrid Systems. *Nanoscale* **2014**, *7*, 4598–4810.

(5) Bonaccorso, F.; Lombardo, A.; Hasan, T.; Sun, Z.; Colombo, L.; Ferrari, A. C. Production and Processing of Graphene and 2d Crystals. *Mater. Today* **2012**, *15*, 564–589.

(6) Mak, K. F.; Lee, C.; Hone, J.; Shan, J.; Heinz, T. F. Atomically Thin $MoS_2$: A New Direct-Gap Semiconductor. *Phys. Rev. Lett.* **2010**, *105*, 136805.

(7) Splendiani, A.; Sun, L.; Zhang, Y.; Li, T.; Kim, J.; Chim, C.-Y.; Galli, G.; Wang, F. Emerging Photoluminescence in Monolayer $MoS_2$. *Nano Lett.* **2010**, *10*, 1271–1275.

(8) Mak, K. F.; He, K.; Shan, J.; Heinz, T. F. Control of Valley Polarization in Monolayer $MoS_2$ by Optical Helicity. *Nat. Nanotechnol.* **2012**, *7*, 494–498.

(9) Zeng, H.; Dai, J.; Yao, W.; Xiao, D.; Cui, X. Valley Polarization in $MoS_2$ Monolayers by Optical Pumping. *Nat. Nanotechnol.* **2012**, *7*, 490–493.

(10) Cao, T.; Wang, G.; Han, W.; Ye, H.; Zhu, C.; Shi, J.; Niu, Q.; Tan, P.; Wang, E.; Liu, B.; Feng, J. Valley-Selective Circular Dichroism of Monolayer Molybdenum Disulphide. *Nat.*

*Commun.* **2012**, *3*, 887.

(11) Ugeda, M. M.; Bradley, A. J.; Shi, S.-F.; da Jornada, F. H.; Zhang, Y.; Qiu, D. Y.; Ruan, W.; Mo, S.-K.; Hussain, Z.; Shen, Z.-X.; Wang, F.; Louie, S. G.; Crommie, M. F. Giant Bandgap Renormalization and Excitonic Effects in a Monolayer Transition Metal Dichalcogenide Semiconductor. *Nat. Mater.* **2014**, *13*, 1091–1095.

(12) Chernikov, A.; Berkelbach, T. C.; Hill, H. M.; Rigosi, A.; Li, Y.; Aslan, O. B.; Reichman, D. R.; Hybertsen, M. S.; Heinz, T. F. Exciton Binding Energy and Nonhydrogenic Rydberg Series in Monolayer $WS_2$. *Phys. Rev. Lett.* **2014**, *113*, 076802.

(13) He, K.; Kumar, N.; Zhao, L.; Wang, Z.; Mak, K. F.; Zhao, H.; Shan, J. Tightly Bound Excitons in Monolayer $WSe_2$. *Phys. Rev. Lett.* **2014**, *113*, 026803.

(14) Mak, K. F.; He, K.; Lee, C.; Lee, G. H.; Hone, J.; Heinz, T. F.; Shan, J. Tightly Bound Trions in Monolayer $MoS_2$. *Nat. Mater.* **2013**, *12*, 207–211.

(15) Jones, A. M.; Yu, H.; Ghimire, N. J.; Wu, S.; Aivazian, G.; Ross, J. S.; Zhao, B.; Yan, J.; Mandrus, D. G.; Xiao, D.; Yao, W.; Xu, X. Optical Generation of Excitonic Valley Coherence in Monolayer $WSe_2$. *Nat. Nanotechnol.* **2013**, *8*, 634–638.

(16) Ross, J. S.; Wu, S.; Yu, H.; Ghimire, N. J.; Jones, A. M.; Aivazian, G.; Yan, J.; Mandrus, D. G.; Xiao, D.; Yao, W.; Xu, X. Electrical Control of Neutral and Charged Excitons in a Monolayer Semiconductor. *Nat. Commun.* **2013**, *4*, 1474.

(17) You, Y.; Zhang, X.-X.; Berkelbach, T. C.; Hybertsen, M. S.; Reichman, D. R.; Heinz, T. F. Observation of Biexcitons in Monolayer $WSe_2$. *Nat. Phys.* **2015**, *11*, 477–481.

(18) Koppens, F. H. L.; Mueller, T.; Avouris, P.; Ferrari, A. C.; Vitiello, M. S.; Polini, M. Photodetectors Based on Graphene, Other Two-Dimensional Materials and Hybrid Systems. *Nat. Nanotechnol.* **2014**, *9*, 780–793.

(19) Ross, J. S.; Klement, P.; Jones, A. M.; Ghimire, N. J.; Yan, J.; Mandrus, D. G.; Taniguchi, T.; Watanabe, K.; Kitamura, K.; Yao, W.; Cobden, D. H.; Xu, X. Electrically Tunable Excitonic Light-Emitting Diodes Based on Monolayer $WSe_2$ P-N Junctions. *Nat. Nanotechnol.* **2014**, *9*, 268–272.

(20) Wang, R.; Ruzicka, B. A.; Kumar, N.; Bellus, M. Z.; Chiu, H.-Y.; Zhao, H. Ultrafast and Spatially Resolved Studies of Charge Carriers in Atomically Thin Molybdenum Disulfide.

*Phys. Rev. B* **2012**, *86*, 45406.

(21) Shi, H.; Yan, R.; Bertolazzi, S.; Brivio, J.; Gao, B. Exciton Dynamics in Suspended Monolayer and Few-Layer MoS$_2$ 2D Crystals. *ACS Nano* **2013**, *7*, 1072–1080.

(22) Sim, S.; Park, J.; Song, J.-G. G.; In, C.; Lee, Y.-S. S.; Kim, H.; Choi, H. Exciton Dynamics in Atomically Thin MoS2: Interexcitonic Interaction and Broadening Kinetics. *Phys. Rev. B* **2013**, *88*, 75434.

(23) Sun, D.; Rao, Y.; Reider, G. A.; Chen, G.; You, Y.; Brézin, L.; Harutyunyan, A. R.; Heinz, T. F. Observation of Rapid Exciton-Exciton Annihilation in Monolayer Molybdenum Disulfide. *Nano Lett.* **2014**, *14*, 5625–5629.

(24) Nie, Z.; Long, R.; Sun, L.; Huang, C.-C.; Zhang, J.; Xiong, Q.; Hewak, D. W.; Shen, Z.; Prezhdo, O. V; Loh, Z.-H. Ultrafast Carrier Thermalization and Cooling Dynamics in Few-Layer MoS$_2$. *ACS Nano* **2014**, *8*, 10931–10940.

(25) Borzda, T.; Gadermaier, C.; Vujicic, N.; Topolovsek, P.; Borovsak, M.; Mertelj, T.; Viola, D.; Manzoni, C.; Pogna, E. A. A.; Brida, D.; Antognazza, M. R.; Scotognella, F.; Lanzani, G.; Cerullo, G.; Mihailovic, D. Charge Photogeneration in Few-Layer MoS$_2$. *Adv. Funct. Mater.* **2015**, *25*, 3351–3358.

(26) Mai, C.; Barrette, A.; Yu, Y.; Semenov, Y. G.; Kim, K. W.; Cao, L.; Gundogdu, K. Many-Body Effects in Valleytronics: Direct Measurement of Valley Lifetimes in Single-Layer MoS$_2$. *Nano Lett.* **2014**, *14*, 202–206.

(27) Kumar, N.; Cui, Q.; Ceballos, F.; He, D.; Wang, Y.; Zhao, H. Exciton-Exciton Annihilation in MoSe$_2$ Monolayers. *Phys. Rev. B* **2014**, *89*, 125427.

(28) Cui, Q.; Ceballos, F.; Kumar, N.; Zhao, H. Transient Absorption Microscopy of Monolayer and Bulk WSe$_2$. *ACS Nano* **2014**, *8*, 2970–2976.

(29) Steinhoff, A.; Rösner, M.; Jahnke, F.; Wehling, T. O.; Gies, C. Influence of Excited Carriers on the Optical and Electronic Properties of MoS$_2$. *Nano Lett.* **2014**, *14*, 3743–3748.

(30) Palummo, M.; Bernardi, M.; Grossman, J. C. Exciton Radiative Lifetimes in Two-Dimensional Transition Metal Dichalcogenides. *Nano Lett.* **2015**, *15*, 2794–2800.

(31) Qiu, D. Y.; da Jornada, F. H.; Louie, S. G. Optical Spectrum of $MoS_2$: Many-Body Effects and Diversity of Exciton States. *Phys. Rev. Lett.* **2013**, *111*, 216805.

(32) Cheiwchanchamnangij, T.; Lambrecht, W. R. L. Quasiparticle Band Structure Calculation of Monolayer, Bilayer, and Bulk $MoS_2$. *Phys. Rev. B* **2012**, *85*, 205302.

(33) Ramasubramaniam, A. Large Excitonic Effects in Monolayers of Molybdenum and Tungsten Dichalcogenides. *Phys. Rev. B* **2012**, *86*, 115409.

(34) Carvalho, A.; Ribeiro, R. M.; Castro Neto, A. H. Band Nesting and the Optical Response of Two-Dimensional Semiconducting Transition Metal Dichalcogenides. *Phys. Rev. B* **2013**, *88*, 115205.

(35) Lee, J.-U.; Park, J.; Son, Y.-W.; Cheong, H. Anomalous Excitonic Resonance Raman Effects in Few-Layered $MoS_2$. *Nanoscale* **2015**, *7*, 3229–3236.

(36) Marini, A. Competition between the Electronic and Phonon–mediated Scattering Channels in the Out–of–equilibrium Carrier Dynamics of Semiconductors: An Ab-Initio Approach. *J. Phys. Conf. Ser.* **2013**, *427*, 012003.

(37) Sangalli, D.; Marini, A. Complete Collisions Approximation to the Kadanoff-Baym Equation: A First-Principles Implementation. *J. Phys. Conf. Ser.* **2015**, *609*, 012006.

(38) Sangalli, D.; Marini, A. Ultra-Fast Carriers Relaxation in Bulk Silicon Following Photo-Excitation with a Short and Polarized Laser Pulse. *Europhys. Lett.* **2015**, *110*, 47004.

(39) Attaccalite, C.; Grüning, M.; Marini, A. Real-Time Approach to the Optical Properties of Solids and Nanostructures: Time-Dependent Bethe-Salpeter Equation. *Phys. Rev. B* **2011**, *84*, 245110.

(40) Marini, A.; Hogan, C.; Grüning, M.; Varsano, D. Yambo: An Ab Initio Tool for Excited State Calculations. *Comput. Phys. Commun.* **2009**, *180*, 1392.

(41) Onida, G.; Reining, L.; Rubio, A. Electronic Excitations: Density-Functional versus Many-Body Green's-Function Approaches. *Rev. Mod. Phys.* **2002**, *74*, 601–659.

(42) Haug, H.; Schmitt-Rink, S. Electron Theory of the Optical Properties of Laser-Excited Semiconductors. *Prog. Quantum Electron.* **1984**, *9*, 3–100.

(43) Reynolds, D. C.; Look, D. C.; Jogai, B. Combined Effects of Screening and Band Gap Renormalization on the Energy of Optical Transitions in ZnO and GaN. *J. Appl. Phys.* **2000**, *88*, 5760.

(44) Stopa, M. Band-Gap Renormalization and Excitonic Binding in T-Shaped Quantum Wires. *Phys. Rev. B* **2001**, *63*, 195312.

(45) Ko, P. J.; Abderrahmane, A.; Thu, T. V.; Ortega, D.; Takamura, T.; Sandhu, A. Laser Power Dependent Optical Properties of Mono- and Few-Layer $MoS_2$. *J. Nanosci. Nanotechnol.* **2015**, *15*, 6843–6846.

(46) Sie, E. J.; Frenzel, A. J.; Lee, Y.-H.; Kong, J.; Gedik, N. Intervalley Biexcitons and Many-Body Effects in Monolayer $MoS_2$. *Phys. Rev. B* **2015**, *92*, 125417.

(47) Chernikov, A.; Ruppert, C.; Hill, H. M.; Rigosi, A. F.; Heinz, T. F. Population Inversion and Giant Bandgap Renormalization in Atomically Thin $WS_2$ Layers. *Nat. Photonics* **2015**, *9*, 466–470.

(48) Lee, C.; Yan, H.; Brus, L. E.; Heinz, T. F.; Hone, J.; Ryu, S. Anomalous Lattice Vibrations of Single- and Few-Layer $MoS_2$. *ACS Nano* **2010**, *4*, 2695–2700.

(49) Zhang, X.; Han, W. P.; Wu, J. B.; Milana, S.; Lu, Y.; Li, Q. Q.; Ferrari, A. C.; Tan, P. H. Raman Spectroscopy of Shear and Layer Breathing Modes in Multilayer $MoS_2$. *Phys. Rev. B* **2013**, *87*, 115413.

(50) Dal Conte, S.; Bottegoni, F.; Pogna, E. A. A.; Ambrogio, S.; Bargigia, I.; D'Andrea, C.; De Fazio, D.; Lombardo, A.; Bruna, M.; Ciccacci, F.; Ferrari, A. C.; Cerullo, G.; Finazzi, M. Valley and Spin Dynamics in Monolayer $MoS_2$. **2015**. arXiv:1502.06817 [cond-mat.mtrl-sci]. arXiv.org e-Print archive. http://arxiv.org/abs/1502.06817 (accessed Oct 15, 2015).

(51) Polli, D.; Brida, D.; Mukamel, S.; Lanzani, G.; Cerullo, G. Effective Temporal Resolution in Pump-Probe Spectroscopy with Strongly Chirped Pulses. *Phys. Rev. A* **2010**, *82*, 053809.

(52) Kovalenko, S. A.; Dobryakov, A. L.; Ruthmann, J.; Ernsting, N. P. Femtosecond Spectroscopy of Condensed Phases with Chirped Supercontinuum Probing. *Phys. Rev. A* **1999**, *59*, 2369–2384.

(53) Wang, H.; Zhang, C.; Rana, F. Ultrafast Dynamics of Defect-Assisted Electron-Hole Recombination in Monolayer MoS$_2$. *Nano Lett.* **2015**, *15*, 339–345.

(54) Radisavljevic, B.; Kis, A. Mobility Engineering and a Metal-Insulator Transition in Monolayer MoS$_2$. *Nat. Mater.* **2013**, *12*, 815–820.

(55) Giannozzi, P.; Baroni, S.; Bonini, N.; Calandra, M.; Car, R.; Cavazzoni, C.; Ceresoli, D.; Chiarotti, G. L.; Cococcioni, M.; Dabo, I.; Dal Corso, A.; de Gironcoli, S.; Fabris, S.; Fratesi, G.; Gebauer, R.; Gerstmann, U.; Gougoussis, C.; Kokalj, A.; Lazzeri, M.; Martin-Samos, L.; *et al.* QUANTUM ESPRESSO: A Modular and Open-Source Software Project for Quantum Simulations of Materials. *J. Phys. Condens. Matter* **2009**, *21*, 395502.

(56) Young, P. A. Lattice Parameter Measurements on Molybdenum Disulphide. *J. Phys. D. Appl. Phys.* **1968**, *1*, 936–938.

(57) Rozzi, C. A.; Varsano, D.; Marini, A.; Gross, E. K. U.; Rubio, A. Exact Coulomb Cutoff Technique for Supercell Calculations. *Phys. Rev. B* **2006**, *73*, 205119.

# SUPPORTING INFORMATION

# Photo-Induced Bandgap Renormalization Governs the Ultrafast Response of Single-Layer MoS2

*Eva A. A. Pogna[1,†], Margherita Marsili[2,†], Domenico De Fazio[3], Stefano Dal Conte[1,4], Cristian Manzoni[1,4], Davide Sangalli[5], Duhee Yoon[3], Antonio Lombardo[3], Andrea C. Ferrari[3], Andrea Marini[5,\*], Giulio Cerullo[1,4,\*], Deborah Prezzi[2,\*]*

[1]Dipartimento di Fisica - Politecnico di Milano, Piazza Leonardo da Vinci 32, I-20133 Milano (IT)

[2]Centro S3, Istituto Nanoscience (NANO) - CNR, via Campi 213/a, I-41125, Modena (IT)

[3]Cambridge Graphene Centre, University of Cambridge, Cambridge CB3 0FA (UK)

[4]Istituto di Fotonica e Nanotecnologie (IFN) - CNR, I-20133 Milano (IT)

[5]Istituto di Struttura della Materia (ISM) - CNR, Via Salaria Km 29.3, Monterotondo Stazione (IT)

† These authors contributed equally to the work.

\* Corresponding authors: (GC) giulio.cerullo@polimi.it; (AM) andrea.marini@cnr.it; (DP) deborah.prezzi@nano.cnr.it

## S1. Transient absorption setup

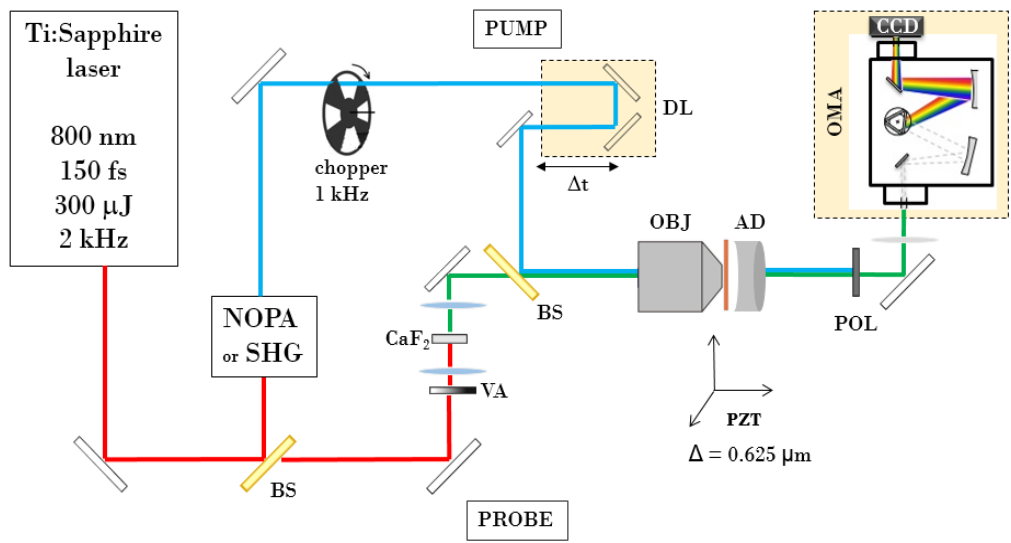

**FigureS1.** Detailed scheme of the pump-probe setup. NOPA: non-collinear optical parametric amplifier; SHG: second-harmonic generation; BS: beam splitter; VA: variable attenuator; OBJ: Cassegrain microscope objective; PZT: piezotranslator; AD: achromatic doublet; OMA: optical multichannel analyzer; POL: linear polarizer; DL: delay line.

## S2. Simulated absorption and exciton composition analysis

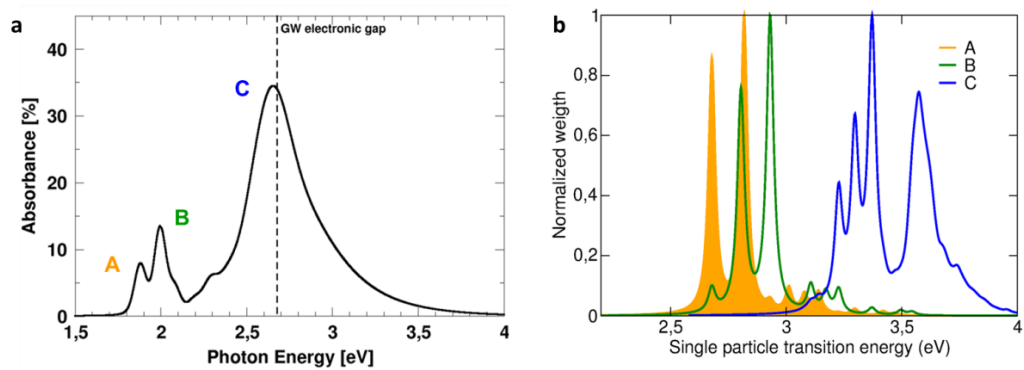

**FigureS2. (a)** Simulated absorbance spectrum of 1L-MoS$_2$, showing the three main excitonic features A, B and C. The dashed line indicates the electronic gap from *GW* calculations. **(b)** The composition of each exciton is reported in terms of weighted single particle transitions.